\documentclass[ 
twocolumn,
showpacs,
preprintnumbers,
bibnotes,
amsmath,
amssymb,
aps,
pra,
superscriptaddress,
longbibliography,
]{revtex4-2}

\usepackage[version=3]{mhchem}
\usepackage[dvipsnames]{xcolor}
\usepackage[colorlinks=false, citecolor=blue, urlcolor=blue, linkcolor=blue]{hyperref}
\usepackage{textcomp}
\usepackage{graphicx}
\usepackage{amsmath}
\usepackage{fixmath}
\usepackage{amsfonts}
\usepackage{amssymb}
\usepackage{braket}
\usepackage{subfigure}
\usepackage{physics}

\begin{document}

\title{The bending rigidity exponent of a two-dimensional crystalline membrane with arbitrary number of flexural phonon modes}
 
\author{D.~A.~Ivanov}
\affiliation{Faculty of Physics, ITMO University, St. Petersburg 197101, Russia}
\email{danil.ivanov@metalab.ifmo.ru}
\author{A.~Kudlis}
\affiliation{Abrikosov Center for Theoretical Physics, MIPT, Dolgoprudnyi, Moscow Region 141701, Russia}
\affiliation{Russian Quantum Center, Skolkovo, Moscow 121205, Russia}
\email{andrewkudlis@gmail.com}
\author{I.~S.~Burmistrov}
\affiliation{L. D. Landau Institute for Theoretical Physics, Semenova 1-a, 142432 Chernogolovka, Russia}
\affiliation{Laboratory for Condensed Matter Physics, HSE University, 101000 Moscow, Russia}
\email{burmi@itp.ac.ru}

\date{\today}

\begin{abstract}
We investigate the elastic behavior of two-dimensional crystalline membrane embedded into  real space taking into account the presence an arbitrary number of flexural phonon modes $d_c$ (the number of out-of-plane deformation field components). The bending rigidity exponent $\eta$ is extracted by numerical simulation via Fourier Monte Carlo technique of the system behaviour in the universal regime. This universal quantity governess the correlation function of out-of-plane deformations at long wavelengths and defines the behaviour of renormalized bending rigidity at small momentum $\varkappa~\sim~1/q^{\eta}$. The resulting numerical estimates of the exponent for various $d_c$ are compared with the numbers obtained from the approximate analytical techniques.
\end{abstract}
\maketitle

\allowdisplaybreaks
Membranes are two-dimensional (2D) surfaces that, depending on external conditions, can be found in different states on the phase diagram. The discovery of graphene \cite{Novoselov2004,Novoselov2005,Zhang2005} and other  atomically thick materials \cite{Novoselov2012} led to renewed interest to 2D crystalline membranes and 
the emergence of new field of flexible 2D materials \cite{2Dmat}. In the case of crystalline membranes, when analyzing their behavior, the presence of non-zero resistance to shear cannot be neglected, which leads to the appearance of effective long-range interaction between out-of-plane deformations. Such an effective interaction, in fact, mediated by the coupling between in-plane and out-of-plane (flexural) deformations, is stiffening the membrane, renormalizing the bending rigidity, and allowing the stabilization of the flat low-temperature phase, within which long-wave fluctuations define a new class of universality~\cite{PhysRevLett.60.2634}. A 2D crystalline membrane in this low-temperature flat phase demonstrates peculiar elastic properties dubbed as {\it anomalous elasticity}. These unusual elastic effects include  crumpling transition controlled by temperature and disorder, power-law scaling of elastic modules with the system size, and consequently, the nonlinear Hooke’s law, negative Poisson ratios, negative thermal expansion coefficient, etc. \cite{Nelson1987,PhysRevLett.60.2634,PhysRevLett.60.2638,David_1988,Aronovitz1989,Guitter1988,Guitter1989,Toner1989,PhysRevLett.69.1209,PhysRevA.45.R2151,Nelson_1991,Radzihovsky1991,PhysRevA.46.1751,Bensimon1992,Radzihovsky1995,Radzihovsky1998}. Currently, there has been significant progress in our theoretical understanding of the anomalous elasticity of crystalline membranes \cite{Kosmrlj:2013,Kosmrlj:2014,Kats2014,Gornyi:2015a,Kats2016,Burmistrov2016,Gornyi2016,Kosmrlj2016,Kosmrlj2017,Doussal2018,Burmistrov2018a,Burmistrov2018b,SAYKIN2020168108,Saykin2020b,Coquand2020,Mauri2020,Mauri2021,LeDoussal2021,Shankar2021,Mauri2022,Metayer2022,Burmistrov2022,Parfenov2022}.

The key property of the low-temperature flat phase is the power law dependence of the renormalized bending rigidity characterized by the value of the so-called bending rigidity exponent, $\eta$. The latter determines the behavior of the correlation function of out-of-plane deformations (OPD) at small momenta. Due to symmetry constraints \cite{Aronovitz1989,Guitter1989}, the remaining exponents can be related with $\eta$ using simple scaling relations. For this reason, the accurate calculation of the magnitude of $\eta$ and verification of the consistency of results obtained by different approaches is an extremely important issue.

Unfortunately, an exact analytical treatment of the bending rigidity exponent is not possible. For this reason various approximate analytical methods for its calculation have been proposed~\cite{PhysRevLett.60.2634,PhysRevLett.60.2638,David_1988,Guitter1989,PhysRevLett.69.1209,PhysRevE.79.040101,Nelson1987,Metayer2022}. The earliest perturbative calculations gave a naive result: $\eta=1$~\cite{Nelson1987}. The recent $4-\varepsilon$-expansion analysis gives the following values of $\eta$: $0.8872$~\cite{Metayer2022} and $0.8670$~\cite{Pikelner_2022}, within the three- and four-loop approximation, respectively. These estimates can be improved by applying various resummation techniques for asymptotic series. In particular, the simplest Pad\'e approximation already gives a value of $0.806$~\cite{Pikelner_2022}. Apart from that, the calculations were also carried out within the nonperturbative renormalization group. For  example, in Ref.~\cite{PhysRevE.79.040101}, the authors extracted  the following estimate: $\eta=0.849$.  

An alternative is the lattice numerical calculations based on the Monte Carlo method. There are various techniques~\cite{G.Gompper_1991,Bowick:1996wz,PhysRevB.80.121405,Troster_2013}. One can work in both real and momentum spaces. The authors believe that one of the most effective methods is the Fourier Monte Carlo (MC) technique. This technique is described in Ref.~\cite{PhysRevB.76.012402}. Its great advantage is that the acceptance rates during the simulations can be adjusted for each wave vector separately. This allows one to reduce the impact of critical slowing down, thereby increasing the accuracy of the final numerical estimates. The authors of this technique have previously analyzed the critical behavior of membranes for the single-component case in their work~\cite{Troster_2013} and extracted the following value: $\eta=0.795(10)$. In this paper, we will use this approach to obtain the results for the multi-component case. In Table~\ref{tab:1}, we summarize numerical values  of $\eta$ obtained for a single-component flexural phonon by various approaches. 

If we assume further that the out-of-plane deformation field can be multi-component (arbitrary number of flexural phonon modes), then the quantity $1/d_c$ can be used as expansion parameter when constructing a perturbation theory. In particular, the $1/d_c$-expansion within first order approximation gives the following series for the exponent: $\eta=2/d_c+\mathcal{O}(1/d_c^2)$~\cite{David_1988}. It is clear that  one should not expect any proper numerical estimates for such a short series when $d_c \approx  1$. Another widespread calculation technique is self-consistent screening approximation (SCSA), which instead sums up a certain subset of perturbation theory diagrams; in pioneering work in this direction, a value of $0.821$ for $d_c=1$ was extracted~\cite{PhysRevLett.69.1209}. The general formula within this approximation reads as follows \cite{PhysRevLett.69.1209}
\begin{align}\label{eqn:scsa}
\eta=\dfrac{4}{d_c+\sqrt{16-2d_c+d_c^2}}.
\end{align}
This non-analytical  expression is still approximate, and the issue of  $1/d_c$-corrections to it is important. Quite recently, the authors in Ref.~\cite{SAYKIN2020168108} managed to extend the results obtained previously in Ref.~\cite{David_1988}, by calculating the next order of perturbation theory. They found the following truncated series:
\begin{align}\label{eqn:second-order}
\eta=\dfrac{2}{d_c} + \dfrac{73-68\zeta(3)}{27}\dfrac{1}{d_c^2} + \mathcal{O}\left(\dfrac{1}{d_c^3}\right).
\end{align}
\begin{table}[t]
    \centering
    \caption{Numerical estimates of  bending rigidity exponents $\eta$ obtained by means of different theoretical approaches and numerical calculations. In the table the following notations are used: MC -- Monte Carlo simulation; MC IP and MC OP stand for Monte Carlo simulations with monitoring mean squared fluctuations of in-plane and out-of-plane phonons, respectively; SCSA -- self-consistent screening-approximation; Sp. cor. -- space correlator; NPRG -- nonperturbative renormalization-group approach; Mol. dyn. -- molecular dynamics; FMC -- Monte Carlo simulations in Fourier space; $\varepsilon$ exp.: 3l -- three-loop $\varepsilon$ expansion calculation; $\varepsilon$ exp.: 4l -- four-loop $\varepsilon$ expansion calculation; [2/2] Pad\'{e} -- result obtained by means of resummation of four-loop $\varepsilon$ series  by means of $[2/2]$ Pad\'{e}- approximant.}
    \label{tab:1}
    \setlength{\tabcolsep}{6.0pt}
    \begin{tabular}{lll|lll}
        \hline
        \hline
        Method & Ref.& $\eta$ & Method & Ref.& $\eta$ \\
        \hline
        MC & \cite{G.Gompper_1991} & $0.60(10)$ & NPRG & \cite{PhysRevE.79.040101} & $0.849$\\
        MC OP & \cite{Bowick:1996wz} & $0.72(4)$ & Mol. dyn. & \cite{PhysRevB.80.121405} & $0.85$\\
        MC IP & \cite{Bowick:1996wz} & $0.750(5)$ & SCSA & \cite{PhysRevLett.69.1209} & $0.821$ \\
        Sp. cor. & \cite{PhysRevLett.116.015901} & $0.78(02)$ & $\varepsilon$ exp.: 3l & \cite{Metayer2022} & $0.8872$\\
        Sp. cor. & \cite{M.Falcioni_1997} & $0.62$ & $\varepsilon$ exp.: 4l & \cite{Pikelner_2022} & $0.8670$\\
        FMC & \cite{Troster_2013}  & $0.795(10)$ & [2/2] Pad\'{e} & \cite{Pikelner_2022} & $0.806$\\
        \hline
        \hline
    \end{tabular}
    \vspace{-1.em}
\end{table}

\noindent In the case of a one-component field, such an expansion gives poor estimates that do not in any way compare with the results of lattice calculations. Paradoxically, the formally incorrect SCSA gives much closer value of $\eta$ that differ by no more than $10$ percent from the results of the MC simulations.

In this Letter we explore the behavior of Green's function of out-of-plane deformations at long wavelengths with respect to a given reference plane governed by the bending rigidity exponent $\eta$. Using the Fourier Monte Carlo technique,  we present the results of a generalization of this method to the case when the number of flexural phonons differs from unity ($d_c>1$). These estimates make it possible to check the convergence of the approximate analytical results that were found previously within the $1/d_c$-expansion method, as well as the SCSA approximation itself.

\noindent\textsf{\color{blue} Model.} We consider a continuous elastic model of a two-dimensional crystalline membrane with $D_{6h}$ or $D_{3h}$ point group embedded in $2+d_c$ dimensional space. Using the Monge representation, the out-of-plane deformations (heights) with respect to a given two-dimensional reference plane with coordinates $\vb{x} = (x_1 , x_2 )$ can be parameterized by the vector function $\vb{h}_{\vb{x}}$, where $d_c$-component structure of height-function corresponds to the presence of multiple flexural (out-of-plane) phonon modes. The main physical meaning is contained in the case $d_c=1$. As was said above, by allowing the height function to be a vector, we test the results of alternative perturbative approaches whose numerical estimates of observables have to be asymptotically correct in $d_c\rightarrow\infty$ limit.

Having integrated the in-plane deformations, in the absence of external stress, the effective energy functional of this model depending only on out-of-plane deformations can be written as follows~\cite{Nelson1987,PhysRevA.45.R2151,PhysRevA.46.1751,Bensimon1992}:
\begin{align}\label{eqn:eff_ham}
    \mathcal{F}_\text{eff}[\vb{h}]\!=\!\frac{\varkappa_0}{2}\!\!\int\!\!{\rm{d}}\vb{x}(\Delta\vb{h})^2+\frac{Y_0}{2}\int\!\!{\rm{d}}\vb{x}\left[P_{ij}^{\perp}K_{ij}\right]^2,
\end{align}
where $Y_0{=}4 \mu_0(\mu_0{+}\lambda_0)/(2\mu_0{+}\lambda_0)$ is the bare magnitude of the Young modulus of 2D crystal and $\varkappa_0$ is the bare bending rigidity. Here $\lambda_0$ and $\mu_0$ are the bare Lam\'e constants. Also we introduced transverse projector $P_{ij}^{\perp}=\delta_{ij}-\partial_i\partial_j/\Delta$, and  auxiliary quadratic quantity $K_{ij}=\partial_i\vb{h}\partial_j\vb{h}/2$. From a field-theoretical point of view, we have an action with an unusual quadratic part that no longer depends on the square, but on the fourth power of momentum, as well as a nonlocal quartic term that describes the effective long-range interaction of out-of-plane deformations induced by their in-plane counterparts. This nonlocality makes the MC simulations within real space  rather complicated. To overcome this difficulties, one can rewrite the effective energy functional (or action) in momentum space. For this purpose, the continuous Fourier amplitudes of height function $\vb{h}_{\vb{x}}$ can be introduced via the following relation
\begin{align}
\vb{h}_{\vb{x}}=\int\limits_{\rm{\Omega}}\!\!\dfrac{\rm{d}\vb{q}}{(2\pi)^2}\vb{h}_{\vb{q}}e^{i\vb{q}\vb{x}}
\end{align}
where the integration space $\rm{\Omega}$ is restricted  by $|\vb{q}|<\Lambda = \pi/a$ and $|\vb{q}|>2\pi/L$. The first constraint corresponds to the condition up to which the continuum model approximation remains valid, or this can be called ultraviolet cutoff (the length $a\simeq \sqrt{\varkappa_0/Y_0}$ is of the order of the lattice spacing). In turn, cutting from below can be interpreted as infrared cutoff, where $L$ stands for the size of the membrane. In terms of Fourier amplitudes the energy functional can be rewritten now in the following form: 
\begin{align}\label{eqn:eff_act_q_space}
\tilde{\mathcal{F}}_\text{eff}[\vb{h}]=\tilde{\mathcal{F}}_\text{eff}^{(2)}[\vb{h}]+\tilde{\mathcal{F}}_\text{eff}^{(4)}[\vb{h}],
\end{align}
where the bending energy is
\begin{align}
    \tilde{\mathcal{F}}_\text{eff}^{(2)}[\vb{h}]&=\dfrac{\varkappa_{0}}{2}\int\dfrac{\dd{\vb{p}}}{(2\pi)^2}\, p^4\left(\vb{h}_{\vb{p}}\vb{h}_{-\vb{p}}\right),\label{eqn:starting_fe}
\end{align}
while the second term corresponds to anharmonic nonlocal part and reads
\begin{multline}
    \tilde{\mathcal{F}}_\textup{eff}^{(4)}[\vb{h}]=\dfrac{Y_{0}}{8}\int\dfrac{\dd\vb{k}}{(2\pi)^2}\dfrac{\dd\vb{k}'}{(2\pi)^2}\int\dfrac{\dd{\vb{q}}}{(2\pi)^2}\left[\vb{k}\times\hat{\vb{q}}\right]^2\left[\vb{k}'\times\hat{\vb{q}}\right]^2\\
    \times\left(\vb{h}_{\vb{k}}\vb{h}_{-\vb{k}-\vb{q}}\right)\left(\vb{h}_{\vb{k}'}\vb{h}_{-\vb{k}'+\vb{q}}\right),\label{eqn:second_term}
\end{multline}
here $\hat{\vb{q}}=\vb{q}/|\vb{q}|$ is normalized momentum vector. 

All the necessary information about the elastic properties of the membrane is contained in the Green’s function, which is determined by the following relation:
\begin{align}
  \langle (\vb{h}_{\vb{k}})_i(\vb{h}_{\vb{-k}})_j \rangle=  \widehat{G}_{ij}(\vb{q})=\delta_{ij}\widehat{G}(\vb{q}).
\end{align}
In the case when there is no shear resistance (liquid membranes), i.e. the shear modulus $\mu$ vanishes, and, as a consequence, the Young’s modulus is also zeroed, the remaining harmonic part of the action~\eqref{eqn:eff_act_q_space} leads to the following behaviour of Green's function:
\begin{align}
   \widehat{G}_0(\vb{q})=\dfrac{T}{\varkappa_0 q^4}.
\end{align}
However, such an approximation is absolutely not suitable for the case of crystalline membranes, whose elastic properties are strongly modified the presence of effective long-range interaction between out-of-plane deformations, which is expressed in the second term of the action~\eqref{eqn:eff_act_q_space}. Indeed, without the second term, the flat phase cannot be stable, which is clearly seen from the expression for the stretching factor (see Ref. \cite{Gornyi:2015a} for a review)
\begin{align}
    \xi^2=
    1 - \langle K_{jj} \rangle & = 
    1-\dfrac{d_c}{2}\int\dfrac{\dd\vb{q}}{(2\pi)^2}q^2 \widehat{G}_0(\vb{q}) \notag \\
    & =1-\dfrac{d_c T}{4\pi\varkappa_0}\ln{\dfrac{L}{2a}},
\end{align}
zeroing of which indicates the transition to the crumpled phase. Physically, the appearing of long-range interaction can be interpretated as stiffening of the membrane at large scales. From the other hand, following the renormalization group terminology, the presence of the second term~\eqref{eqn:second_term} leads to a renormalization of the bare bending rigidity $\varkappa_0$ and Young's modulus, which become a momentum-dependent \cite{PhysRevLett.60.2634,David_1988}: $\varkappa\sim q^{-\eta}$ and $Y_0 \sim q^{2-2\eta}$, where $\eta$ is bending rigidity exponent. This in turn changes the behavior of the Green's function, $\hat{G}(\vb{q}) \sim |\vb{q}|^{\eta-4}$ and also changes the equation of state $\xi^2-1 \sim  d_c T/\varkappa_0 \eta$, thus stabilizing the flat phase at low temperatures. Thus, the numerical value of quantity $\eta$ is key in the theory. 

Having obtained the basic understanding of the model, below we present a discrete analogue of the action~\eqref{eqn:eff_act_q_space}, which is used in specific calculations, as well as some technical aspects of the formalism used earlier in ~\cite{Troster_2013} for $d_c=1$ and generalized in our work.

\noindent\textsf{\color{blue} Technical details.} 
We consider a two-dimensional square\ lattice $\vb{\Omega}$ of linear dimension $ a N \times a N$ with periodic boundary conditions, where $a$ is lattice  constant and $N$ is integer characterizing the number of lattice periods of the sample. In order to employ the Fourier transform we define vector product by its value in the first Brillouin zone and continue it from there by periodicity. We introduce correspondence between the representations of a microstate of the system by the set of real field-values $\vb{h}_{\vb{x}}$ and one given by the collection of Fourier amplitudes $\tilde{\vb{h}}_{\vb{q}}$ with the same physical dimension($\vb{h}_{\vb{q}}=a^2\tilde{\vb{h}}_{\vb{q}}$) as the discretized field $\vb{h}_{\vb{x}}$:
\begin{align}
    \vb{h}_{\mathbf{x}}=\dfrac{1}{N^{2}}\sum\limits_{\vb{q}\in\widetilde{\vb{\Omega}}_q}\tilde{\vb{h}}_{\vb{q}}e^{i\vb{qx}}, &\;\;\tilde{\vb{h}}_{\vb{q}}=\sum\limits_{\vb{x}\in\vb{\Omega}}\vb{h}_{\vb{x}}e^{-i\vb{qx}}.
\end{align}
These asymmetric conventions allow for easy translation of continuum formulae. For the simulation, the  expression~\eqref{eqn:eff_act_q_space} is rewritten in the following discrete dimensionless form \cite{Troster_2013,Saykin2019}:
\begin{multline}
    \tilde{\mathcal{F}}_\text{eff,d}[\tilde{\vb{h}}] =\sum\limits_{\vb{m\neq 0}}\Biggl\{\dfrac{1}{2}{\rm\Delta}_{s,\vb{m}}^2\left|\tilde{\vb{h}}_{\vb{m}}\right|^2+\dfrac{2\pi}{3}\frac{p^2_8}{N^2}\left|S_{\vb{m}}\right|^2\Biggr\},\label{eq:tFF}
\end{multline}
with summation over $\vb{m}=\{m_1,m_2\}$, where $m_{1,2}~=~1,\dots, N$, and Laplacian operator defined as: 
\begin{align}
    {\rm \Delta}_s(\vb{m})=4\left[\sin^2\left(\dfrac{\pi m_1}{N}\right) + \sin^2\left(\dfrac{\pi m_2}{N}\right)\right]^2.
\end{align}
Here and further, we replace all momenta $q_i=2\pi m_i/N$ by the $\sin$-based functions which is connected with representation of derivatives via Fourier transforms of nearest neighbor finite difference operators (see p. 7 in Ref.~\cite{Troster_2013}). In the same way, the function $S_{\vb{m}}$ is introduced
\begin{align}
    S_{\vb{m}}=\sum\limits_{\vb{n}\neq 0}p(\vb{n},\vb{m})
    \tilde{\vb{h}}_{\vb{n}} \tilde{\vb{h}}_{\vb{n}+\vb{m}},
\end{align}
where the summation is performed over $\vb{n}=\{n_1,n_2\}$, $n_{1,2} = 1,\dots, N$, using the  auxiliary function $p(\vb{n},\vb{m})$ which replaces the square of cross product $\left[\vb{k}\times\hat{\vb{q}}\right]^2$ and reads 
\begin{align}
 &   p(\vb{n},\vb{m})=\\
 &   =\dfrac{\left[\sin\! \left(\!\dfrac{2\pi n_1}{N}\!\right)\! \cos\! \left(\!\dfrac{2 \pi m_2}{N}\!\right)\!-\!\cos\!\left(\!\dfrac{2 \pi n_2}{N}\!\right) \sin\!\left(\! \dfrac{2 \pi m_1}{N}\!\right)\!\right]^2}{4 \left[\sin^2\left(\dfrac{\pi m_1}{N}\right) + \sin^2\left(\dfrac{\pi m_2}{N}\right)\right]}.\nonumber
\end{align}

The quantity $p_8$ contains information about interaction strength measured in dimensionless units. 
Such a constant can be roughly estimated as $p_8~=~a\sqrt{3Y_0k_{\textup{B}}T/16\pi\varkappa_0^2}~=~a q_*/\sqrt{2}$, where we introduced a \textit{Ginzburg wave vector} $q_{*}$, which serves as a boundary point, passing through which a crossover occurs from mean-field behavior (where phonon interaction can be neglected) to the critical one with a non-zero exponent $\eta$. In particular, for graphene the relevant parameters are as follows: lattice constant $a$~=~2.46~\textup{~\AA}, $Y_0~\simeq$~22~eV$\cdot$\textup{\AA}$^{-2}$ and $\varkappa_0~\simeq$~1.1~eV. For $T$~=~300 K it gives~$p_8$~=~0.41. 

\begin{figure}[t!]
    \centering
    \includegraphics[width=1\linewidth]{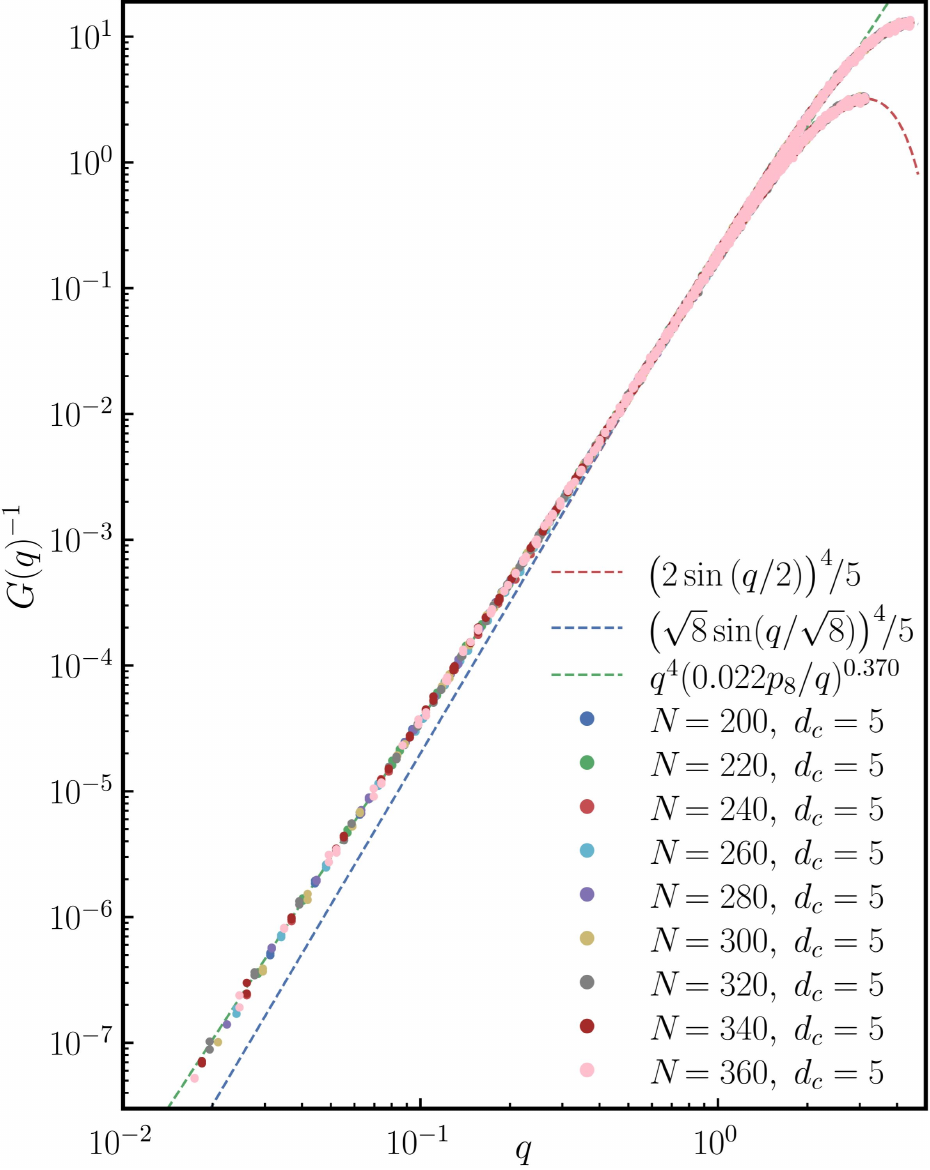}
    \vspace{-1.1em}
    \caption{The log–log plot of inverse Green's function $\widehat{G}(2\pi\vb{q}/N)^{-1}$ for different lattice sizes~$N=200,\;220,\;\ldots,\;360$ with~$d_c~=~5$. The multi-colored data points obtained from the FMC simulations refer to different lattice sizes $N$. The plot shows three analytical functions on a log-log scale: two of them correspond to the initial distribution of the height vector $\tilde{\vb{h}}$ (red and blue dashed lines), and the last one represents linear fit in the area close to zero (green dashed line), the slope of which is the calculated $\eta$.}
    \vspace{-1.em}
    \label{fig:green}
\end{figure}
Everything starts with an initial distribution for the height field $\tilde{\vb{h}}_{\vb{k}}$, which is chosen in the same way for all components. Usually the following dependency is taken as the initial one: $h_{i,\vb{k}} \sim  1/k^2$, for $i=1,\dots,d_c$. After initialization, the Monte Carlo step is calculated, which can be divided into two stages: calculating the lattice energy functional, or rather its change $\Delta \mathcal{F}$, and applying the Metropolis--Hastings (MH) algorithm for the energy functional under shift. The Fourier Monte Carlo (FMC) moves correspond to the following change:
\begin{align}
    \tilde{\vb{h}}_{\vb{q}} \mapsto \tilde{\vb{h}}_{\vb{q}}+a(\vb{q}_0,d_c) \vb{z} \delta_{\vb{q},\vb{q}_0} +a(-\vb{q}_0,d_c)_0\Bar{\vb{z}}\delta_{\vb{q},\vb{-q}_0}\label{eqn:fmc},
\end{align}
where $\vb{z}$ is a random complex vector, which each random component has a bounded modulus, $a(\vb{q},d_c)$ is a phase exploration $\vb{q}$-dependent step, and the wave vector $\vb{q}_0\in\tilde{\vb{\Omega}}$ chosen at random. For each specific $d_c$, we choose a real-valued $a(\vb{q}_0,d_c)$ by hands so that acceptance rate varies between $30$ -- $60$\%. Let us provide additional comment here. As the number of components $d_c$ increases, in the case of uniform and $d_c$-independent choice for function $a=a_0$ the acceptance rate rapidly drops to zero, which is a manifestation of the phenomenon of critical slowing down. It is easy to understand when looking at the individual acceptance rates for moves of type~\eqref{eqn:fmc} for different wave vectors $\vb{q}_0$. One can see that this behavior is a result of the fact that with a uniform choice of the shift, the acceptance coefficients for “small” $\vb{q}_0$-vectors are close to 100\%, whereas for large momenta, which however make up the vast majority of $\vb{q}$-vectors, they can fall well below 30\%. For this reason, it is necessary to choose different values of $a(\vb{q},d_c)$ for different $\vb{q}_0$, which, however, does not violate the detailed balance, which in turn frees the FMC algorithm from critical slowing down~\cite{Troster_2013}. Despite this, the calculation of MC steps itself becomes more difficult as the number of components of the height vector increases, which, combined with a decrease in the numerical value of the critical exponent as $d_c$ grows, leads to an increase in the relative error of calculating $\eta$.
\begin{figure}[b!]
    \centering
    \vspace{-0.3em}
    \includegraphics[width=1\linewidth]{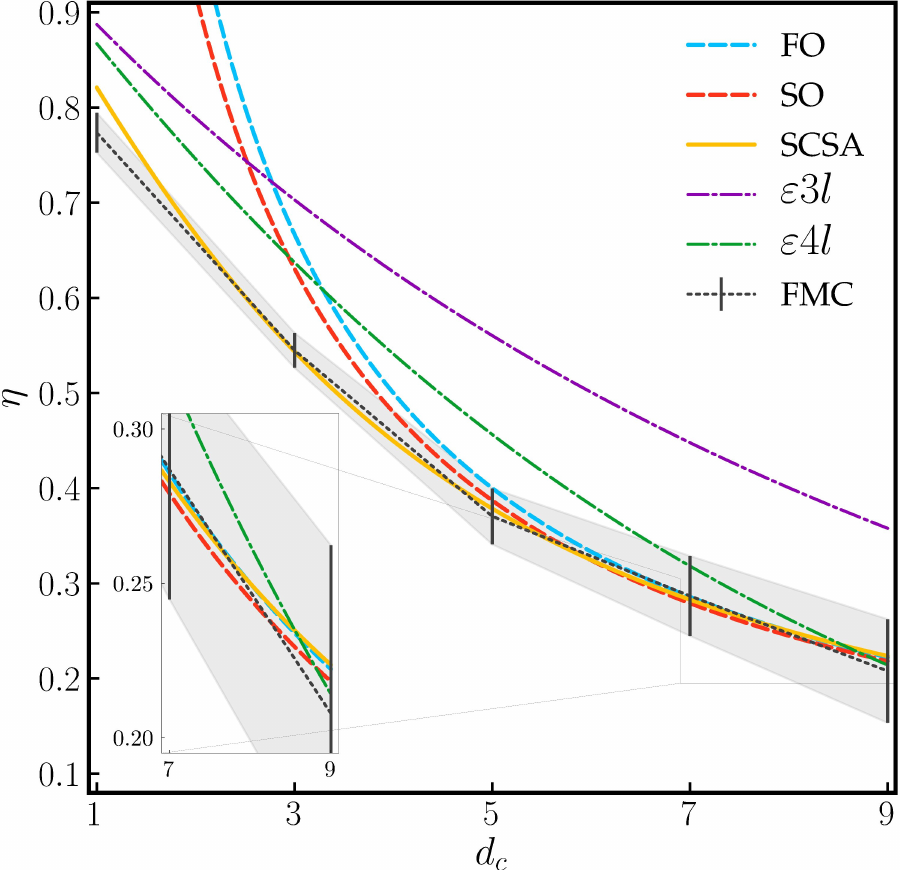}
    \vspace{-2.3em}
    \caption{Dependencies of $\eta$ on the number of components of the height vector obtained within the first-order (FO) in 1/$d_c$ (blue dashed line), and within the second-order (SO) in 1/$d_c$~\eqref{eqn:second-order} (red dashed line), by self-consistent screening approximation (SCSA)~\eqref{eqn:scsa} (yellow solid line), by three-loop $\varepsilon$ expansion ($\varepsilon 3l$) (purple dash-dotted line), by four-loop $\varepsilon$ expansion ($\varepsilon 4l$) (green dash-dotted line) and by means of FMC approach with applied errors (gray vertical lines at points corresponding to odd natural values $d_c= 1, 3, 5, 7, 9$, linearly connected by a dotted line).}
    \vspace{-1.em}
    \label{fig:flux_dep}
\end{figure}

Thus, having completed the MC step~\eqref{eqn:fmc}, we determine the magnitude of the energy change and compare $-\Delta \mathcal{F}$ with the logarithm of a uniformly distributed random number $R \in [0,1]$. If the condition of the Metropolis-Hastings algorithm works for the resulting change ($-\Delta \mathcal{F} > \ln{R}$), then the energy functional changes to a new one, and the configuration of the height field $\tilde{\vb{h}}_{\vb{q}_0}$ shifts by $a(\vb{q}_0,d_c)\vb{z}$, and $\tilde{\vb{h}}_{-\vb{q}_0}$ is shifted by $a(-\vb{q}_0,d_c)\overline{\vb{z}}$.

After thermalization and performing a certain number of Monte Carlo steps, we find the average value for the Green's function for each component of the height vector:
\begin{align}
    G_i(\vb{q})=\langle |h_{i,\vb{q}}|^2\rangle_{\textup{MC}}\sim \dfrac{1}{q^{4-\eta_i}},
\end{align}
where each $i$ maps the component ($i = 1, 2, \ldots, d_c$). Putting points from all $\ln\left(1/G_i(\vb{q})\right)$ on same figure, we should fit the data
by the following expression $c+~\!\!(4-~\!\!\eta)\ln\left(q\right)$. 

\begin{table}[t]
 \centering
    \caption{The results of FMC simulations for $\eta$ for different number of components $d_c$. Let us add values of $\eta$ for $d_c=~\!\!5,7,9$ computed via first-order within $1/d_c$ (FO), second-order within $1/d_c$ (SO), self-consistent screening approximations (SCSA), three-loop $\varepsilon$ expansion ($\varepsilon 3l$) and four-loop $\varepsilon$ expansion ($\varepsilon 4l$). "TW" stands for "this work".}
    \label{tab:2}
     \setlength{\tabcolsep}{9.3pt}
    \begin{tabular}{lllll}
      \hline
      \hline
      &Ref.&$d_c=5$&$d_c=7$&$d_c=9$\\
      \hline
        $\eta$&TW&$0.370(29)$&$0.287(42)$&$0.208(55)$\\
        $\eta_{\textup{FO}}$&\cite{David_1988}&$0.4$&$0.286$&$0.222$\\
        $\eta_{\textup{SO}}$&\cite{SAYKIN2020168108}&$0.387$&$0.279$&$0.218$\\
        $\eta_{\textup{SCSA}}$&\cite{PhysRevLett.69.1209}&$0.379$&$0.283$&$0.224$\\
        $\eta_{\varepsilon 3l}$&\cite{Metayer2022}&$0.560$&$0.448$&$0.358$\\
        $\eta_{\varepsilon 4l}$&\cite{Pikelner_2022}&$0.456$&$0.318$&$0.214$\\        
      \hline
    \end{tabular}
    \vspace{-1.em}
\end{table}

Numerical experiments were carried out for different lattice sizes $N$~=~200,~220,~$\ldots$,~360 and for different number of components $d_c$~=~1,~3,~5,~7,~9. For completeness, we present the behavior of the Green's function on a double logarithmic scale in the Fig.~\ref{fig:green}. For convenience, we plot a linear fit, on the basis of which the values of the critical exponent $\eta$ are extracted.

\noindent\textsf{\color{blue}  Results and discussion.} Our main result for $\eta$ is shown in the form of a trend in the Fig.~\ref{fig:flux_dep}, and through the numbers presented in a separate Table~\ref{tab:2}. In this figure, for comparison, we depict the results of the $1/d_c$ expansion in both the first and second orders of perturbation theory~\eqref{eqn:second-order}; in addition, the trends obtained based on the SCSA formula~\eqref{eqn:scsa} and $\varepsilon$ expansions within three- and four-loop approximations are plotted. First of all, Fig.~\ref{fig:flux_dep} shows quite an important feature --  all the methods begin to converge to each other as $d_c$ increases, at least up to the observed value $d_c=9$ (although for three loops in $\varepsilon$ this is not so obvious). Moreover, one can see that SCSA and $1/d_c$ within both orders begin to fully coincide in the vicinity of $d_c=7$, but for $d_c=9$ they begin to diverge slightly and, as a result, SCSA turns out to be higher than first-order and second-order approximations in $1/d_c$. As for the $\varepsilon$ expansion, its convergence is not so fast, but also noticeable. On the other hand, for small values of $d_c$ ($d_c \lesssim 3$), we can conclude that only formally incorrect SCSA technique and $\varepsilon$ expansion (in both orders) pass the strength test, relying on our numerical calculation, within which for small $d_c$ we can guarantee a small error. It would be interesting to see whether the critical exponent will continue to pursuit closer to SCSA or whether there will be a turning point when the result begins to press closer to the first- and second-order trends in $1/d_c$. However, this check is beyond current technical capabilities, since it will require very high accuracy of calculations during very long simulations.

\noindent\textsf{\color{blue}  Conclusion.} 
In our study we successfully determine the bending rigidity exponent $\eta$ for two-dimensional crystalline membranes embedded in a real space, considering various numbers of flexural phonon modes. The numerical simulations conducted through Fourier Monte Carlo technique reveal the universal behavior of the system, offering insights into the correlation function of out-of-plane deformations and the renormalized bending rigidity at small momentum scales. By comparing the numerical estimates with approximate analytical techniques, the Letter provides a comprehensive understanding of the elastic behavior of such membranes, shedding light on their physical properties in different scenarios. Based on the results, it can be seen that the $1/d_c$-expansion begins to work well somewhere around $d_c=5$. Surprisingly, the SCSA result~\eqref{eqn:scsa}, which has no formal analytical justification except the limit $d_c\to \infty$, gives very reasonable estimate for numerical value of $\eta$ for $d_c\leqslant 5$.

As a further development of this work we will apply our simulations to the model with a random curvature \cite{PhysRevA.45.R2151,PhysRevA.46.1751,Bensimon1992} in order to detect the finite $T$ transition between clean flat phase and rippled flat phase predicted in Ref. \cite{Saykin2020b} within $1/d_c$ expansion and in Ref. \cite{PhysRevE.103.L031001} by means of two-loop $\varepsilon$ expansion for $4-\varepsilon$ dimensional membrane. 

Other direction in which numerical methods used in our work can be applied is numerical calculation of the differential and absolute Poisson's ratios. Although, their values are predicted by means of perturbation theory at $d_c\gg1$ \cite{Burmistrov2018a,SAYKIN2020168108}, their dependence on $d_c$ at small $d_c$ is not known.

Finally, our numerical scheme could be generalized to the case of a crystalline membrane with orthorhombic crystal symmetry.  It would be possible to verify the prediction of Ref. \cite{Burmistrov2022} that in this case of reduced symmetry the renormalization of the bending rigidity tensor is still controlled by the exponent $\eta$.

\noindent\textsf{\color{blue}  Acknowledgement}. We are grateful to David Saykin for the hints relative to Monte Carlo calculations in the one-component case. We acknowledge the computing time provided to us at computer facilities at Landau Institute. The work of I.S.B. was funded in part by Russian Ministry of Science and Higher Education (project FFWR-2024-0015) as well as by Basic research program of HSE. The work of A.K. is supported by the Ministry of Science and Higher Education of the Russian Federation (Goszadaniye, project No. FSMG-2023-0011).

\appendix

\bibliography{lit.bib}

\end{document}